\documentclass[aps,pra,twocolumn,10pt,nofootinbib,longbibliography,floatfix]{revtex4-2}
\usepackage{amsmath,amssymb,bm,graphicx,placeins}
\usepackage[dvipsnames]{xcolor}
\usepackage{hyperref,orcidlink}
\hypersetup{colorlinks=true,linkcolor=BrickRed,citecolor=RoyalBlue,urlcolor=RoyalBlue}
\graphicspath{{figures/}}
\newcommand{\ket}[1]{|#1\rangle}
\newcommand{\bra}[1]{\langle#1|}
\newcommand{\Tr}{\operatorname{Tr}}
\newcommand{\E}{\mathbb E}
\newcommand{\MMD}{\operatorname{MMD}}

\begin{document}
\title{Generative quantum extreme learning machines}
\author{Marcin P\l{}odzie\'n\,\orcidlink{0000-0002-0835-1644}}
\affiliation{Instytut Fizyki Teoretycznej, Wydzia\l{} Fizyki, Astronomii i Informatyki Stosowanej,
Uniwersytet Jagiello\'nski, \L{}ojasiewicza 11, PL-30-348 Krak\'ow, Poland}
\date{\today}
\begin{abstract}
Training the circuit of a quantum generative model is hampered by barren plateaus, where gradients vanish exponentially with the number of qubits. In the present work we propose a generative protocol based on the paradigm of quantum extreme learning machines, in which the circuit is never trained. A random latent vector is encoded into a quantum state, evolved with a fixed unitary and measured, and only a linear output layer is trained. The encoding of the latent variable determines the functions available to the model, while the unitary decides which combinations of them are seen by the trained layer. With the help of Pauli propagation we show that for a Clifford unitary each feature is a single product of sines and cosines of the encoded angles, so that a Clifford circuit, even a strongly entangling one, does not create new functions. Doping the circuit with $T$ gates mixes these products and changes the nonstabilizerness of the state, but the space of functions remains the one fixed by the encoding. For brickworks of Haar-random two-qubit gates the features are no longer single products, and their variation with the latent variable decreases with the depth of the circuit. For the proposed encoding and for one- and two-dimensional distributions, generated both from expectation values and from single measurements, neither the entanglement produced by a Clifford reservoir nor the additional nonstabilizerness of Haar-random brickworks systematically improves the generator, and a reservoir with no gates at all is competitive with entangling circuits.
\end{abstract}
\maketitle

\section{Introduction}
\label{sec:intro}
Generative modeling is among the most extensively studied applications of near-term quantum devices~\cite{biamonte2017quantum,cerezo2021variational,benedetti2019generative,gao2018quantum}. In the quantum circuit Born machine a parametrized state is measured in the computational basis, and the Born rule provides a probability distribution over the exponentially many bitstrings~\cite{liu2018differentiable,cheng2018information,coyle2020born}. In quantum generative adversarial networks, on the other hand, a quantum generator is trained against a discriminator, which can be classical or quantum~\cite{dallaire2018quantum,zoufal2019quantum}. In both cases the quantum circuit itself is subject to optimization. In consequence, each training step requires estimates from the quantum device, and, moreover, expressive circuits suffer from the barren plateau problem, where gradients vanish exponentially with the number of qubits~\cite{mcclean2018barren,cerezo2021cost,holmes2022connecting,ragone2024,rudolph2024trainability}.

A cheaper way is to keep the entangling part of the circuit fixed. In previous works we considered a Born machine with a fixed scrambling layer, where only single-qubit rotations are trained~\cite{plodzien2026qsbm}, and we found that a Clifford entangler, which adds no nonstabilizerness, reaches the loss of a near-Haar-random one once the circuit is deep enough~\cite{plodzien2026qbm}. Quantum extreme learning machines and quantum reservoir computers go one step further, namely the quantum system is not trained at all~\cite{fujii2017harnessing,fujii2021quantum,mujal2021opportunities,innocenti2023potential}. The input is encoded into a quantum state, which then evolves under a fixed unitary, a few observables are measured and a linear readout is fitted to them. Such machines have been applied to time series~\cite{mujal2023time,franceschetto2026backaction}, to image classification~\cite{delorenzis2025harnessing} and, recently, to data generation~\cite{ferreira2025levels}. For rotation encodings the measured features are trigonometric polynomials~\cite{schuld2021effect}, whose frequencies and number limit the functions available to the readout~\cite{xiong2025fundamental}. Gross and Rieser describe these features through the Pauli transfer matrix and note that a Clifford reservoir only permutes them, with possible sign changes~\cite{gross2026qelm}.

Much less is known about the quantum resources which are actually needed by the generated distribution. Entanglement is generated by the multi-qubit gates, and quantum correlations can enlarge the class of distributions which a generative model is able to represent~\cite{gao2022enhancing}. Nonstabilizerness, quantified for example by the stabilizer R\'enyi entropy (SRE)~\cite{leone2022stabilizer,haug2023stabilizer}, is the resource separating universal quantum computation from stabilizer circuits, the latter being efficiently simulable~\cite{gottesman1998heisenberg,aaronson2004,veitch2014,howard2017}. Its role has been investigated in variational algorithms~\cite{spriggs2024quantumresources}. Xia et al. found it necessary for quantum reservoir computing with qubit resetting~\cite{xia2026}, while Ivaki et al. studied how conditional-$T$ doping affects quantum resources and performance on temporal tasks~\cite{ivaki2026optimal}. The role of entanglement in quantum extreme learning machines has been examined in Ref.~\cite{delorenzis2025entanglement}. However, in the case of an extreme learning machine generating data, it is still not clear what the reservoir actually adds to the output distribution.

Here, we address this question first for Clifford reservoirs, where the answer can be given exactly, and then for brickworks of Haar-random two-qubit gates, which approximate a Haar-random unitary with growing depth. A Clifford circuit maps each Pauli string onto a single string, so that the measured observables can be propagated backward at a polynomial cost~\cite{gottesman1998heisenberg,aaronson2004,rudolph2026pauli}, and it entangles the encoded qubits without adding nonstabilizerness~\cite{leone2022stabilizer}. Doping with $T$ gates introduces non-Clifford operations~\cite{leone2021quantum,haferkamp2023efficient}, while the propagation remains possible at a cost exponential only in the number of $T$ gates~\cite{bravyi2016}.

We show that each feature of a Clifford reservoir is a single product of sines and cosines of the encoded angles, so that the reservoir creates no new functions and only selects the products that reach the trained head. The reservoir without gates is thus a classical control, and for the considered one- and two-dimensional targets it is competitive with entangling Clifford circuits, both when the head reads expectation values and when it reads single measurements. Brickworks of Haar-random gates make the expectation-value fits worse at larger depths. Clifford reservoirs appear in Ref.~\cite{gross2026qelm} as a limiting case of Hamiltonian reservoirs; here we apply their structure to generation with a dyadic latent encoding and derive the bitstring weights of the single-shot mode.

\begin{figure*}[t]
\includegraphics[width=\textwidth]{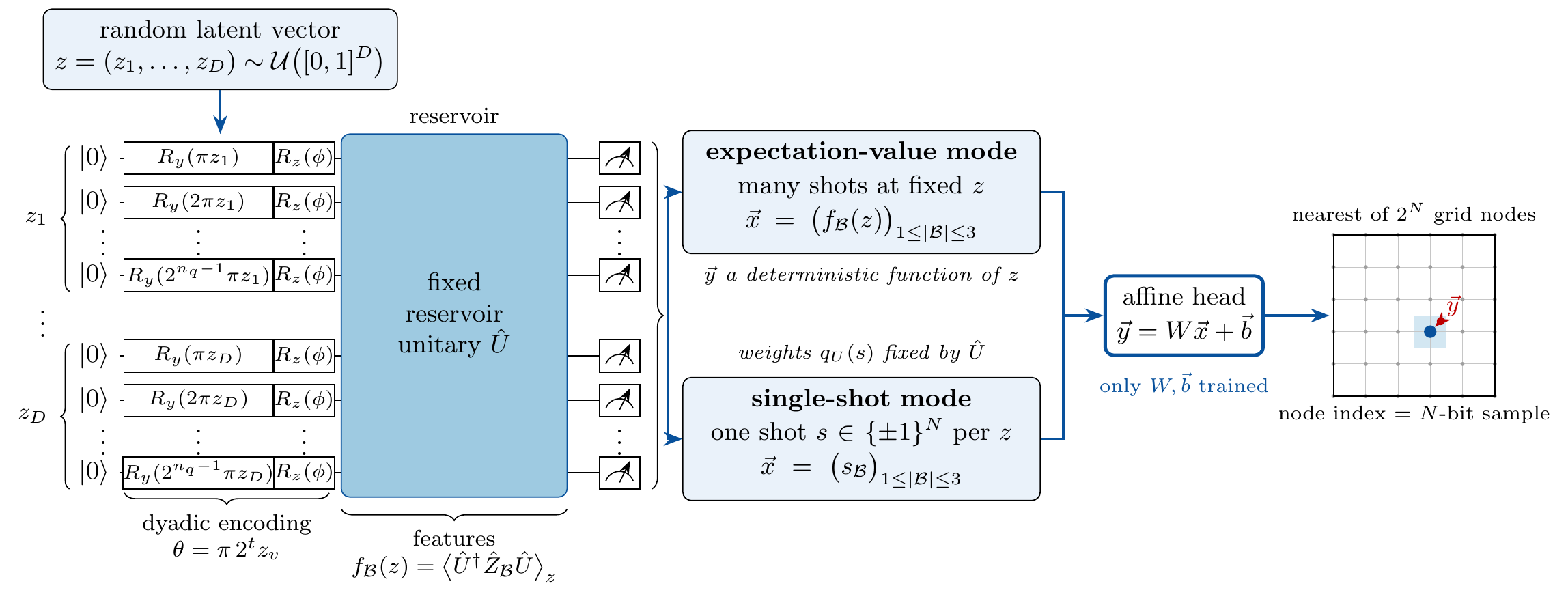}
\caption{Schematic of the quantum extreme learning machine as a generator. A random latent vector $z=(z_1,\dots,z_D)$ is drawn uniformly from $[0,1]^D$, where $D$ equals the dimension $d$ of the target distribution. Each coordinate $z_v$ is encoded on $n_q=N/D$ qubits by the rotations $R_y(\pi2^tz_v)$ with $t=0,\dots,n_q-1$, each followed by $R_z(\phi)$, so that each further qubit turns twice as fast, and vertical dots stand for the qubits and coordinates not drawn. The encoded state passes through a fixed reservoir unitary $\hat U$, a Clifford circuit or a random brickwork of Haar-random two-qubit gates, and is measured in the computational basis. Here $f_{\mathcal B}(z)$ is the mean of the measured parity $s_{\mathcal B}$ at fixed $z$. The bias $\vec b$ replaces the empty-support feature. In the expectation-value mode the head receives these features, $\vec x=\vec f(z)$, and the generated sample $\vec y$ is a deterministic function of $z$. In the single-shot mode it receives the parities of one measured bitstring, $\vec x=\vec x(s)$, and $\hat U$ enters only through the bitstring weights $q_U(s)$. Only $W$ and $\vec b$ are trained. The nearest grid node to $\vec y=W\vec x+\vec b$ supplies the $N$-bit output index.}
\label{fig:protocol}
\end{figure*}

\section{Pauli string propagation}
\label{sec:pauli}
A Pauli string on $N$ qubits is a tensor product $\hat P=\bigotimes_{i=1}^{N}\hat p_i$ with $\hat p_i\in\{\hat I,\hat X,\hat Y,\hat Z\}$. The sites with $\hat p_i\ne\hat I$ form its support; their number is its weight. Let us consider a product state with single-qubit Bloch vectors $\bm r_i$. In such a state the expectation value of a string factorizes,
\begin{equation}
 \big\langle\hat P\big\rangle=\prod_{i=1}^{N}r_i^{p_i},
 \label{eq:product-general}
\end{equation}
where $r_i^{p_i}$ is the Bloch component selected by the letter $\hat p_i$, and $r_i^{I}=1$.

For a state evolved by a unitary $\hat U$, the expectation value of $\hat P$ equals that of the Heisenberg-evolved operator $\hat U^\dagger\hat P\hat U$ in the initial state. For a generic $\hat U$ this operator is a superposition of exponentially many strings. Clifford unitaries are exactly those that map every Pauli string to a single string~\cite{gottesman1998heisenberg,aaronson2004},
\begin{equation}
 \hat U^\dagger\hat P\hat U=\varepsilon\,\hat Q,\qquad\varepsilon=\pm1,
 \label{eq:pullback}
\end{equation}
and we call $\varepsilon\hat Q$ the pull-back of $\hat P$. The Clifford group is generated by the Hadamard gate $\hat H$, the phase gate $\hat S$ and the CNOT gate, and the pull-back of any string follows from the images of single-qubit operators listed in Table~\ref{tab:images}. For a circuit $\hat U=\hat G_n\cdots\hat G_1$ we conjugate $\hat P$ with the gates one after another in the reverse order, i.e., starting from the last gate. For example, for the cluster layer $\hat U=\hat H^{\otimes N}\prod_i\mathrm{CZ}_{i,i+1}$ we obtain
\begin{equation}
 \hat U^\dagger\hat Z_i\hat U=\hat Z_{i-1}\hat X_i\hat Z_{i+1}
 \label{eq:cluster-pullback}
\end{equation}
for every interior qubit, which is the stabilizer generator of the one-dimensional cluster state. In other words, measuring one qubit after the cluster layer amounts to measuring a product of Bloch components on three neighboring qubits, where the circuit determines which component is contributed by each of them.

\begin{table}[t]
\caption{Heisenberg images $\hat G^\dagger\hat P\hat G$ of single-qubit Pauli operators under the gates used in this work. $\mathrm{CNOT}_{i,j}$ has control $i$ and target $j$, $\hat S=\mathrm{diag}(1,\mathrm i)$ and $\hat T=\mathrm{diag}(1,e^{\mathrm i\pi/4})$; $\hat Y=\mathrm i\hat X\hat Z$.}
\label{tab:images}
\begin{ruledtabular}
\begin{tabular}{lcccc}
gate & $\hat X_i$ & $\hat Z_i$ & $\hat X_j$ & $\hat Z_j$\\\hline
$\hat H_i$ & $\hat Z_i$ & $\hat X_i$ & & \\
$\hat S_i$ & $-\hat Y_i$ & $\hat Z_i$ & & \\
$\hat T_i$ & $(\hat X_i-\hat Y_i)/\sqrt2$ & $\hat Z_i$ & & \\
$\mathrm{CNOT}_{i,j}$ & $\hat X_i\hat X_j$ & $\hat Z_i$ & $\hat X_j$ & $\hat Z_i\hat Z_j$\\
$\mathrm{CZ}_{i,j}$ & $\hat X_i\hat Z_j$ & $\hat Z_i$ & $\hat Z_i\hat X_j$ & $\hat Z_j$\\
\end{tabular}
\end{ruledtabular}
\end{table}

A binary-vector representation allows a Pauli string to pass through $n$ Clifford gates in $O(n)$ local updates~\cite{aaronson2004}. Evaluating the product in Eq.~\eqref{eq:product-general} then takes $O(N)$ operations. No state vector is needed. This is the simplest case of Pauli propagation~\cite{rudolph2026pauli}, which also estimates expectation values of typical random circuits efficiently~\cite{angrisani2024classically} and simulates quantum convolutional networks on standard benchmarks~\cite{bermejo2024quantum}; its relation to the trainability of variational models is discussed in Ref.~\cite{cerezo2023does}.

For the $T$ gate, $\hat T^\dagger\hat Z\hat T=\hat Z$, while
\begin{equation}
 \hat T^\dagger\hat X\hat T=\frac{\hat X-\hat Y}{\sqrt2},\qquad
 \hat T^\dagger\hat Y\hat T=\frac{\hat X+\hat Y}{\sqrt2}.
 \label{eq:T}
\end{equation}
A string that meets $n_T$ $T$ gates on transverse letters therefore branches into at most $2^{n_T}$ strings, each of which we propagate through the remaining Clifford gates as before. The cost thus grows exponentially only in the number of $T$ gates met by the string~\cite{rudolph2026pauli}, in line with the cost of simulating Clifford circuits with few $T$ gates~\cite{bravyi2016}.

\section{Quantum extreme learning machine as a generator}
\label{sec:model}
\subsection{Generative protocol}
\label{sec:protocol}
Our goal is to generate samples from a target distribution on a grid in $d$ dimensions. The generative protocol, sketched in Fig.~\ref{fig:protocol}, starts with drawing a latent vector $z=(z_1,\dots,z_D)$ of the same dimension, $D=d$, uniformly from $[0,1]^D$. Next, we encode the latent vector into $N$ qubits through local rotations, $n_q=N/D$ of them for every latent coordinate, with $D$ a divisor of $N$. We denote by $v(i)\in\{1,\dots,D\}$ the latent coordinate carried by qubit $i$ and by $t_i\in\{0,\dots,n_q-1\}$ its dyadic frequency index, so that the state of qubit $i$ reads
\begin{equation}
 \ket{\psi_i(z)}=\hat R_z(\phi)\hat R_y(\theta_i)\ket0,\qquad
 \theta_i=\pi\,2^{t_i}z_{v(i)},
 \label{eq:encoding}
\end{equation}
with a fixed azimuth $\phi=\pi/4$. The encoded state is a product state with Bloch vectors $\bm r_i=(\cos\phi\sin\theta_i,\sin\phi\sin\theta_i,\cos\theta_i)$. As $z_v$ runs over $[0,1]$, the qubit with $t_i=0$ turns through half a circle, and every further qubit of the same coordinate turns twice as fast. The encoding therefore resolves the coordinate at all dyadic scales, much like the binary digits of $z_v$. Exponentially growing encoding weights generate an exponentially large frequency spectrum from few encoding gates~\cite{shin2023exponential}, and our dyadic choice is a variant of this idea.

The encoded state then passes through a fixed unitary $\hat U$, the reservoir, which is chosen once and never trained. After the reservoir all qubits are measured in the computational basis. A measurement returns a bitstring $s\in\{\pm1\}^N$, and from it we form the parities $s_{\mathcal B}=\prod_{i\in\mathcal B}s_i$, which are the measured values of the $Z$ strings $\hat Z_{\mathcal B}=\prod_{i\in\mathcal B}\hat Z_i$. We keep all supports with $1\le|\mathcal B|\le3$, i.e., $M=N+\binom N2+\binom N3$ strings, which commute and are therefore obtained from the same bitstrings. The quantum device passes on only these $M$ numbers, which we collect in a vector $\vec x\in\mathbb R^M$, to a trained affine head that maps them to a point
\begin{equation}
 \vec y=W\vec x+\vec b
 \label{eq:generator}
\end{equation}
of the target space $\mathbb R^d$. Finally, we assign $\vec y$ to the nearest node of a fixed grid of $2^N$ nodes, and the index of this node, written in binary, is the $N$-bit sample produced by the generator. The matrix $W\in\mathbb R^{d\times M}$ and the vector $\vec b\in\mathbb R^d$ are the only trained parameters.

There are two ways to obtain $\vec x$ from the measurements. We can either average the parities over many shots of the same encoded state, or take them from a single shot, and these two possibilities define the two modes discussed in Secs.~\ref{sec:expectation} and \ref{sec:singleshot}, respectively. Both modes rely on the same Pauli propagation, which we discuss first.

\subsection{Features and bitstring probabilities}
When $\hat U$ is a Clifford circuit, every $Z$ string has a pull-back $\varepsilon_{\mathcal B}\hat Q_{\mathcal B}$, and Eq.~\eqref{eq:product-general} gives its expectation value in the encoded state,
\begin{equation}
 \begin{split}
 f_{\mathcal B}(z)&=\big\langle\hat U^\dagger\hat Z_{\mathcal B}\hat U\big\rangle_z\\
 &=\varepsilon_{\mathcal B}\prod_{i=1}^{N}g_{q_{\mathcal B i}}\big(\theta_i(z)\big),
 \end{split}
 \label{eq:product}
\end{equation}
where $q_{\mathcal Bi}$ is the letter of $\hat Q_{\mathcal B}$ on qubit $i$, $g_I=1$, $g_Z(\theta)=\cos\theta$, $g_X(\theta)=\cos\phi\sin\theta$ and $g_Y(\theta)=\sin\phi\sin\theta$. We call $f_{\mathcal B}$ the feature of the support $\mathcal B$. Each feature is thus one product of sines and cosines of the encoded angles, known before any data are generated. This is the Clifford case of the Pauli-transfer-matrix description of Ref.~\cite{gross2026qelm}. Without a reservoir, $\hat Z_{\mathcal B}$ returns the product of $\cos\theta_i$ over $i\in\mathcal B$, whereas with the cluster layer the measured $\hat Z_i$ returns
\begin{equation}
 f_{\{i\}}(z)=\cos\phi\,\cos\theta_{i-1}\sin\theta_i\cos\theta_{i+1}.
 \label{eq:cluster}
\end{equation}
The reservoir thus spreads a measured string over more qubits and exchanges cosines for sines.

The same features fix the whole distribution of the measured bitstring. The projector onto a bitstring can be written as
\begin{equation}
 \ket s\bra s=\prod_{i=1}^{N}\frac{1+s_i\hat Z_i}{2}=\frac{1}{2^N}\sum_{\mathcal B}s_{\mathcal B}\hat Z_{\mathcal B},
 \label{eq:projector}
\end{equation}
with the sum over all $2^N$ subsets of qubits, and the Born probability of $s$ at fixed $z$ becomes
\begin{equation}
 p_U(s\,|\,z)=2^{-N}\sum_{\mathcal B}s_{\mathcal B}\,f_{\mathcal B}(z).
 \label{eq:born}
\end{equation}
Equation~\eqref{eq:born} is the Walsh expansion of $p_U(s\,|\,z)$, with coefficients $f_{\mathcal B}(z)$. For a Clifford reservoir, each coefficient is given by the product in Eq.~\eqref{eq:product}.

\subsection{Expectation-value mode}
\label{sec:expectation}
In the expectation-value mode we average the measured parities over many shots at fixed $z$ and pass the resulting vector $\vec f(z)=(f_{\mathcal B}(z))_{1\le|\mathcal B|\le3}$ to the head. Its output $\vec y=W\vec f(z)+\vec b$ is a smooth map of the latent domain. We use $D=d$.

At $0<\phi<\pi/2$ an $X$ and a $Y$ letter give proportional functions, while at either endpoint one of them vanishes; any interior value gives the same model after whitening, and we take $\phi=\pi/4$. A Clifford reservoir thus selects at most $M$ of the $3^N$ products of $1$, $\cos\theta_i$ and $\sin\theta_i$, and different reservoirs generally select different, non-nested sets. Two measured strings whose images differ only by signs or by $X\leftrightarrow Y$ give the same feature, so that some reservoirs select fewer than $M$ distinct products; among the random brickworks used below this happens for four of the ten.

A circuit of CZ and $Z$ gates commutes with every measured string and returns exactly the features obtained without gates, however much it entangles the encoded state.

\subsection{Single-shot mode}
\label{sec:singleshot}
In the single-shot mode a draw of $z$ is followed by one measurement, giving $s$ with probability $p_U(s\,|\,z)$. The head maps its parities $\vec x(s)=(s_{\mathcal B})_{1\le|\mathcal B|\le3}$ to $W\vec x(s)+\vec b$. Repeated draws of the same bitstring give the same output, regardless of $z$, so its weight is $q_U(s)=\int p_U(s\,|\,z)\,dz$. Using Eq.~\eqref{eq:born},
\begin{equation}
 \begin{split}
 \nu&=\sum_{s}q_U(s)\,\delta_{W\vec x(s)+\vec b},\\
 q_U(s)&=2^{-N}\sum_{\mathcal B}s_{\mathcal B}\int f_{\mathcal B}(z)\,dz .
 \end{split}
 \label{eq:discrete}
\end{equation}
Equation~\eqref{eq:discrete} assigns a fixed weight to each of $2^N$ bitstring-labelled points. Training changes their positions through the shared parameters $W$ and $\vec b$. The resulting distribution can have separated support, although these positions cannot be chosen independently. After quantization, several points may occupy one grid node while others remain empty.

The weights require the latent means of all $2^N$ parities, including those outside the readout. For Clifford reservoirs these means are analytic (App.~\ref{app:qu}). Without gates, every nonconstant feature is a product of cosines $\cos(\pi2^tz_v)$ with zero mean, giving $q_I(s)=2^{-N}$.

A Clifford reservoir exchanges cosines for sines. A product can have a nonzero mean only if, for every latent coordinate it depends on, it contains an odd number of sines and a factor on the qubit with $t_i=0$ (App.~\ref{app:qu}). Such products make the weights nonuniform. Without gates, the head acts on parities of $N$ fair coin tosses. A Clifford reservoir replaces the uniform bitstring law by the weights $q_U(s)$, computed before training.

\begin{figure*}[t]
\includegraphics[width=\textwidth]{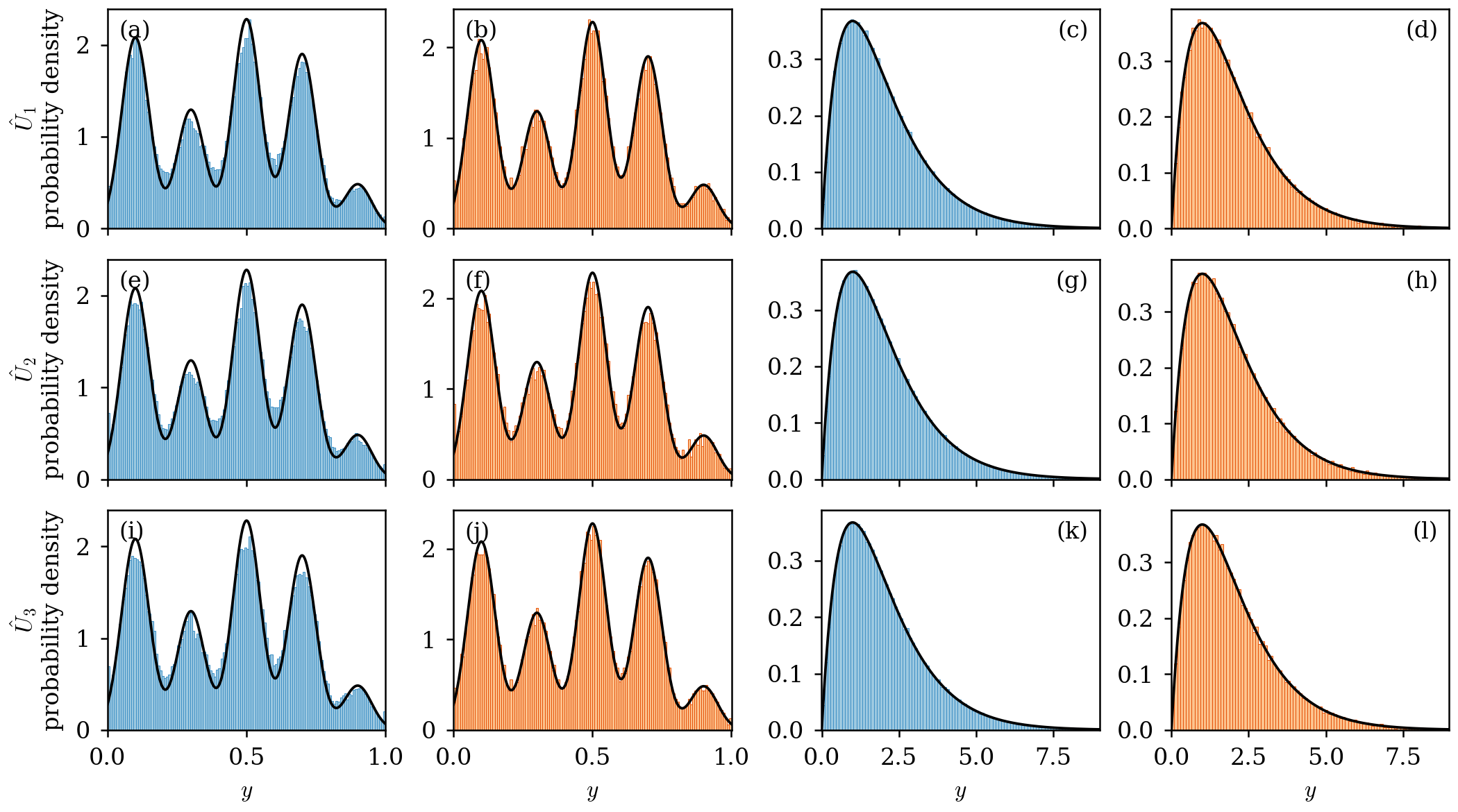}
\caption{Generated distributions for the five-peak Gaussian mixture (first two columns, bin index rescaled to $[0,1]$) and the Gamma distribution (last two columns), for the reservoirs $\hat U_1=\openone$ (top row), the cluster layer $\hat U_2$ (middle row) and the random Clifford brickwork $\hat U_3$ (bottom row). Blue bars show the expectation-value mode and orange bars the single-shot mode. The bars give the probability density of the generated distribution on the $4096$-node grid, grouped into $128$ bins, and the black lines show the probability density of the target. The squared maximum mean discrepancy of Eq.~\eqref{eq:mmd}, evaluated in the training coordinates of Appendix~\ref{app:methods}, is $6.2\times10^{-4}$ in (a), $8.0\times10^{-5}$ in (b), $5.6\times10^{-8}$ in (c), $3.1\times10^{-7}$ in (d), $8.5\times10^{-4}$ in (e), $3.6\times10^{-4}$ in (f), $6.5\times10^{-8}$ in (g), $4.5\times10^{-7}$ in (h), $9.4\times10^{-4}$ in (i), $5.3\times10^{-5}$ in (j), $1.4\times10^{-7}$ in (k) and $4.8\times10^{-7}$ in (l). The Gamma panels are cut at $y=9$. In all panels $N=12$ and $M=298$.}
\label{fig:gallery1d}
\end{figure*}

\subsection{Classical content of the generator}
\label{sec:classical}
In the following, we compare the two modes with classical models for Clifford reservoirs. In the expectation-value mode, the latent variable is the only source of randomness, and the head acts on $M$ fixed trigonometric functions of it, which is a linear model on fixed features, as in random-feature methods~\cite{rahimi2007random} and their quantum counterpart, quantum kitchen sinks~\cite{wilson2018quantum}. Since Eq.~\eqref{eq:product} gives the features in closed form at a cost polynomial in $N$, a quantum device would only estimate known functions.

In the single-shot mode, the randomness comes from the measurement, and the head acts on the parities of a random bitstring. The generated distribution depends on the device only through the weights $q_U(s)$, which we compute exactly from the latent means of the features. Without a reservoir these weights are those of $N$ fair coins. Sampling the output bits of a general Clifford circuit on a non-stabilizer product state is hard classically unless the polynomial hierarchy collapses~\cite{jozsa2014classical}, although not for the shallow nearest-neighbor circuits used here~\cite{jozsa2006simulation}. This result concerns the bitstrings at a fixed input, i.e., the distribution $p_U(s\,|\,z)$, whereas the generator uses only its latent average $q_U(s)$, whose classical complexity for general Clifford circuits we do not address. The single-shot mode is therefore the only place where a device could contribute beyond a classical computation, and the comparison with the identity reservoir in Sec.~\ref{sec:results} tests whether this contribution matters.

\begin{figure*}[t]
\includegraphics[width=\textwidth]{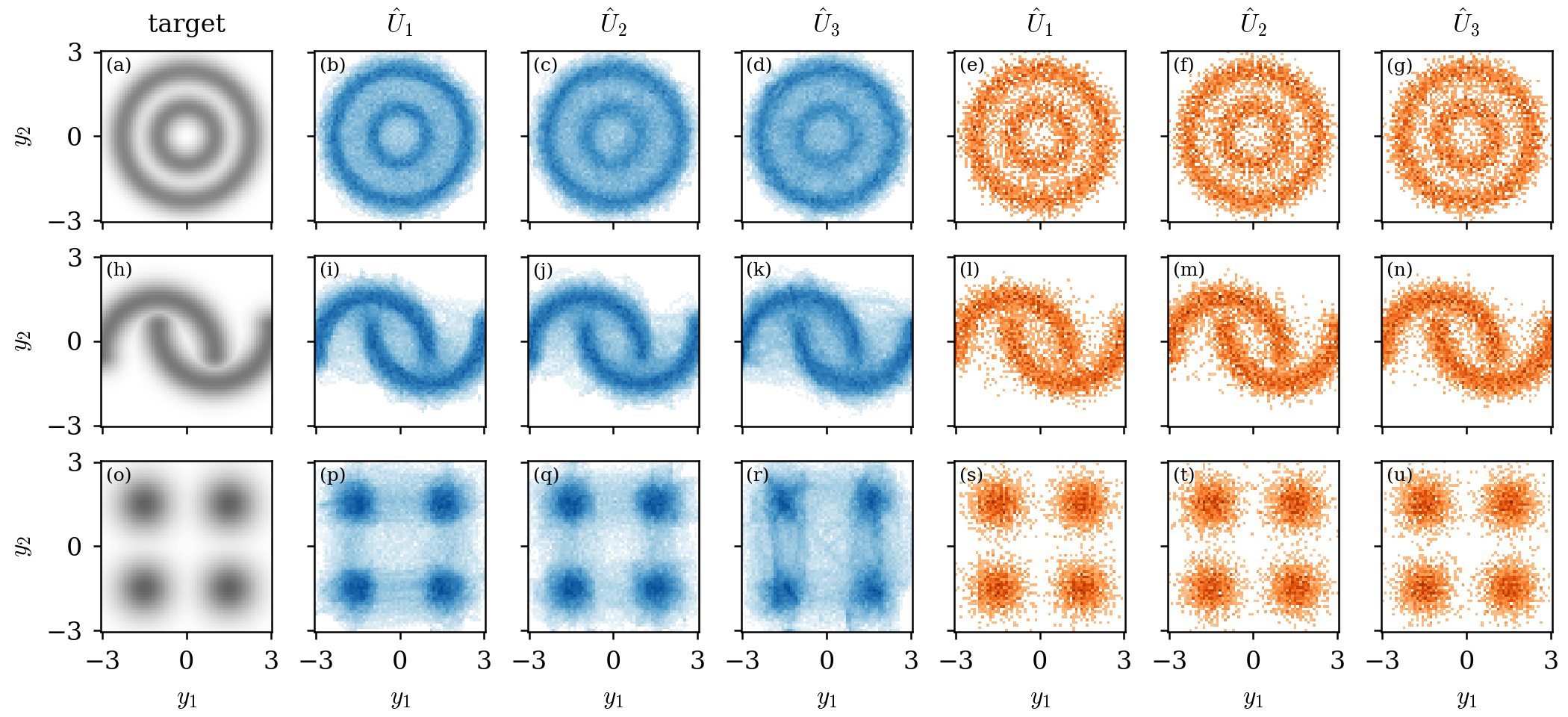}
\caption{Two-dimensional targets (gray, first column) and generated distributions on the $64\times64$ grid for two rings (top row), two moons (middle row) and four lobes (bottom row). Blue panels show the expectation-value mode and orange panels the single-shot mode, each for the reservoirs $\hat U_1=\openone$, $\hat U_2$ (cluster layer) and $\hat U_3$ (random Clifford brickwork). In each row the color coordinate is $\sqrt{p/p_{\max}}$, from white at $p=0$ to the darkest color at the largest probability $p_{\max}$ in the row, shared by all seven panels. Every pixel shows the probability of a single grid node, and no smoothing is applied. The expectation-value distributions are histograms built from $2^{17}$ latent points, i.e., about $32$ points per node, therefore part of the visible fluctuations comes from this finite sample. The squared maximum mean discrepancy of Eq.~\eqref{eq:mmd}, evaluated in the training coordinates of Appendix~\ref{app:methods}, is $3.4\times10^{-4}$ in (b), $6.6\times10^{-4}$ in (c), $8.2\times10^{-4}$ in (d), $3.3\times10^{-4}$ in (e), $2.8\times10^{-4}$ in (f), $2.8\times10^{-4}$ in (g), $1.1\times10^{-4}$ in (i), $1.9\times10^{-4}$ in (j), $6.5\times10^{-4}$ in (k), $4.0\times10^{-4}$ in (l), $1.9\times10^{-4}$ in (m), $1.6\times10^{-4}$ in (n), $1.0\times10^{-3}$ in (p), $4.3\times10^{-4}$ in (q), $6.6\times10^{-3}$ in (r), $6.9\times10^{-5}$ in (s), $1.5\times10^{-4}$ in (t) and $6.9\times10^{-5}$ in (u). In all panels $N=12$ and $n_q=6$ per coordinate.}
\label{fig:gallery2d}
\end{figure*}

\subsection{Training and targets}
\label{sec:training}
Each target is a probability distribution $p_j$ on the $2^N$ nodes $\vec y_j$ of the output grid. In both modes the generator gives a set of points with fixed weights, namely the head outputs $\vec y_a=W\vec x_a+\vec b$ with weights $w_a$. In the expectation-value mode the points come from $n$ latent points $z_a$ with equal weights $w_a=1/n$, and in the single-shot mode from the $2^N$ bitstrings $s_a$ with weights $w_a=q_U(s_a)$. We train the head by minimizing the squared Wasserstein distance between this weighted set of points and the target~\cite{peyre2019computational},
\begin{equation}
 W_2^2=\min_{\pi}\sum_{a,j}\pi_{aj}\,\big|\vec y_a-\vec y_j\big|^2,
 \label{eq:w2}
\end{equation}
where the minimum runs over all couplings $\pi_{aj}\ge0$ with marginals $\sum_j\pi_{aj}=w_a$ and $\sum_a\pi_{aj}=p_j$. In one dimension the optimal coupling pairs the points and the nodes in sorted order, and we compute $W_2^2$ exactly. In two dimensions we use the sliced distance~\cite{bonneel2015sliced}
\begin{equation}
 \mathcal L=\frac1J\sum_{k=0}^{J-1}W_2^2\big(\nu_{\vec e_k},p_{\vec e_k}\big),\qquad
 \vec e_k=(\cos\alpha_k,\sin\alpha_k),
 \label{eq:sliced}
\end{equation}
with $J=32$ fixed directions $\alpha_k=(k+\tfrac12)\pi/J$, where $\nu_{\vec e}$ and $p_{\vec e}$ are the one-dimensional distributions of the projections $\vec e\cdot\vec y$ of the generated points and of the target nodes. We evaluate the loss on the continuous head outputs and apply the assignment to grid nodes only after training. Only the head enters the loss, so that no derivative of the quantum circuit is needed.

To assess a trained generator, we assign the generated points to their nearest grid nodes and assess the resulting distribution $\nu_j$ by its squared maximum mean discrepancy to the target~\cite{gretton2012kernel},
\begin{equation}
 \MMD^2=\sum_{j,j'}(\nu_j-p_j)(\nu_{j'}-p_{j'})\,\kappa(\vec y_j,\vec y_{j'}),
 \label{eq:mmd}
\end{equation}
where the kernel $\kappa(\vec y,\vec y')=\sum_\sigma e^{-|\vec y-\vec y'|^2/2\sigma^2}$ is a sum of Gaussians of widths $\sigma\in\{0.2,0.5,1,2,4\}$. It vanishes only when the two distributions coincide, and we use it for evaluation only, not in training.

In the single-shot mode the loss is a sum over all $2^N$ points, with analytic weights for Clifford reservoirs and weights from numerical quadrature for the Haar brickworks of Sec.~\ref{sec:haar}, which limits this classical training to moderate $N$. In the expectation-value mode the loss approximates the integral over the latent space by $n$ latent points, and on a device every such point requires estimates of all $M$ expectation values. Every head starts from a generic random initialization that matches only the mean and covariance of the target (App.~\ref{app:methods}).

We use $N=12$ qubits, i.e., $n_q=12$ for one-dimensional targets on $4096$ nodes and $n_q=6$ per coordinate for two-dimensional targets on a $64\times64$ grid, with $M=298$ features. The one-dimensional targets are the five-peak Gaussian mixture with unequal weights of Ref.~\cite{plodzien2026qbm}, with means $\mu_j=(j-\tfrac12)2^N/5$ and common width $2^N/20$ on the bin index, and the Gamma distribution $p(y)\propto y\,e^{-y}$, evaluated at $4096$ equally spaced nodes on $[0,12]$. The two-dimensional targets are five of the six targets of Ref.~\cite{plodzien2026qbm}, all but the spirals, on $[-3,3]^2$, namely two concentric rings, two interleaving moons, a diagonal cross, three diagonal stripes and four Gaussian lobes, all but the last concentrated around thin, curved or diagonal ridges. Their definitions are collected in App.~\ref{app:methods}.

In the following we compare three reservoirs, namely $\hat U_1=\openone$, i.e., no gates, the cluster layer $\hat U_2=\hat H^{\otimes N}\prod_i\mathrm{CZ}_{i,i+1}$ of Eq.~\eqref{eq:cluster-pullback}, and a random Clifford brickwork $\hat U_3$ on an open chain of qubits $0,\dots,N-1$. In a brickwork of depth $L=4$ the first and third layers apply two-qubit gates to the pairs $(0,1),(2,3),\dots,(N-2,N-1)$, while the second and fourth layers apply them to the shifted pairs $(1,2),(3,4),\dots,(N-3,N-2)$. For $N=12$ this gives $22$ gates in total. Each gate is drawn uniformly from the $11520$ elements of the two-qubit Clifford group, up to a global phase. We draw $\hat U_3$ once, before any training, and keep it fixed for all targets. To test how typical $\hat U_3$ is, we draw nine further random brickworks in the same way (Sec.~\ref{sec:results}). All reservoirs are used in both modes.

\section{Generated distributions}
\label{sec:results}
All results are obtained for $N=12$ qubits. For one-dimensional targets all twelve qubits encode the single latent coordinate, at the frequencies $\pi2^t$ with $t=0,\dots,11$, while for two-dimensional targets six qubits encode each coordinate, with $t=0,\dots,5$. With $M=298$ measured parities the head has $d(M+1)$ coefficients, i.e., $299$ in one and $598$ in two dimensions.

\subsection{One-dimensional targets}
Figure~\ref{fig:gallery1d} shows the generated distributions for the three reservoirs $\hat U_1$, $\hat U_2$ and $\hat U_3$. For the five-peak mixture the generator recovers in both modes the unequal heights of the peaks as well as the dips between them, and for the Gamma distribution the sharp rise at the origin together with the long tail. In the expectation-value mode the peaks are somewhat broader than those of the target, and a small part of the probability is put at the edges of the grid. For the five peaks the single-shot output has the smaller discrepancy, even though it consists of only $4096$ points with fixed weights, while its total variation distance is larger because of the fluctuations from node to node. The three reservoirs give practically the same picture. In the expectation-value mode $\hat U_1$, which returns only products of cosines, even fits the five peaks better than the two entangling circuits $\hat U_2$ and $\hat U_3$.

Compared with ten random brickworks, $\hat U_1$ is the best for the five-peak mixture and second best for the Gamma distribution with expectation values, and second best and at the median, respectively, in the single-shot mode. $\hat U_3$ is the best of the ten for the five-peak mixture, so that Fig.~\ref{fig:gallery1d} shows a favorable brickwork, and a typical one broadens the peaks more.

\subsection{Two-dimensional targets}
\label{sec:2d}
The rings, the moons and the four lobes remain recognizable for all three reservoirs in Fig.~\ref{fig:gallery2d}, and the two readouts differ mainly in the low-density regions. With expectation values, probability extends across the gap between the rings and between the lobes. The single-shot lobes are better separated, but the probability varies sharply from node to node, since only $4096$ weighted points are placed on a grid of $4096$ nodes. The smooth kernel of Eq.~\eqref{eq:mmd} is only weakly sensitive to these fluctuations, which show up instead in a larger total variation distance. The diagonal cross and the three stripes, not shown, behave similarly.

The differences between reservoirs can reach an order of magnitude, and no reservoir is the best for all targets. In the expectation-value mode $\hat U_1$ gives the lowest discrepancy for the rings, $\hat U_2$ is the best for the lobes, and $\hat U_3$ is the weakest of the three. In the single-shot mode, on the other hand, $\hat U_1$ is, among the identity and the ten random brickworks, the weakest for the rings and the stripes and the tenth for the cross, although the head receives the same parities for every reservoir and only their weights $q_U(s)$ depend on it. The weights of $\hat U_3$ are almost uniform, and evaluating its trained heads under the uniform weights of $\hat U_1$, without retraining, practically does not change the discrepancy for any of the five targets. The single-shot differences between $\hat U_3$ and $\hat U_1$ thus come mainly from the training, which ends in different minima of the loss, rather than from the weights.

\section{Doping with \texorpdfstring{$T$}{T} gates}
\label{sec:doping}
In the following, we consider Clifford reservoirs doped with $T$ gates, which we treat analytically without training doped generators. Nonstabilizerness, or magic, is what separates efficiently simulable Clifford dynamics from universal quantum computation~\cite{gottesman1998heisenberg,aaronson2004,bravyi2005UniversalQuantumComputation}. In fault-tolerant architectures it is supplied by magic-state distillation~\cite{bravyi2005UniversalQuantumComputation,veitch2014,howard2017}. It is a resource distinct from entanglement~\cite{fux2024entanglement} and has been studied in many-body states~\cite{liu2022many,white2021,oliviero2022ising,tarabunga2023manybody,turkeshi2023pauli}, random circuits~\cite{niroula2023phase,Turkeshi2025}, variational circuits, neural-network states and generative models~\cite{spriggs2024quantumresources,sinibaldi2025nonstabilizerness,lipardi2026magic,kruger2026generative,plodzien2026qbm}. We quantify it by the stabilizer R\'enyi entropy (SRE)~\cite{leone2022stabilizer,haug2023stabilizer,tarabunga2023manybody,sierant2026computing},
\begin{equation}
 M_2(\ket\psi)=-\log_2\bigg[\frac{1}{2^N}\sum_{\hat P}\bra\psi\hat P\ket\psi^4\bigg],
 \label{eq:sre}
\end{equation}
where the sum runs over all $4^N$ Pauli strings. The SRE is non-negative, it vanishes if and only if $\ket\psi$ is a stabilizer state, it does not change under Clifford unitaries, and it approaches $N-2$ for typical Haar-random states~\cite{leone2022stabilizer,turkeshi2023pauli}. A Clifford reservoir therefore adds no nonstabilizerness to the encoded state, which already carries some through the single-qubit rotations of Eq.~\eqref{eq:encoding}. A $T$ gate can raise or lower the SRE, depending on the state on which it acts, so that the number of $T$ gates does not fix the nonstabilizerness of the reservoir output.

By Eq.~\eqref{eq:T}, a $T$ gate leaves a $Z$ letter unchanged but turns a transverse letter into a sum of two. A measured string that meets $n_T$ $T$ gates on transverse letters therefore branches into at most $2^{n_T}$ strings $\hat Q_k$, and its feature becomes the fixed combination
\begin{equation}
 f(z)=\sum_kc_k\,f_{\hat Q_k}(z),
 \label{eq:doped}
\end{equation}
where $f_{\hat Q_k}$ is the product of Eq.~\eqref{eq:product-general} with the letters of $\hat Q_k$. The smallest example has two qubits. In the reservoir $\hat U=\hat H_0\hat T_0\,\mathrm{CNOT}_{1,0}$, a CNOT gate with control on qubit $1$ and target on qubit $0$ is followed by a $T$ gate and a Hadamard gate on qubit $0$, and we measure $\hat Z_0$. Propagating backward, the Hadamard gate turns $\hat Z_0$ into $\hat X_0$ and the $T$ gate splits $\hat X_0$ into $(\hat X_0-\hat Y_0)/\sqrt2$. The CNOT gate then leaves $\hat X_0$ unchanged but attaches $\hat Z_1$ to $\hat Y_0$, so that
\begin{equation}
 \hat U^\dagger\hat Z_0\hat U=\frac{\hat X_0-\hat Y_0\hat Z_1}{\sqrt2}.
 \label{eq:two-qubit}
\end{equation}
The two branches end on different strings, and the feature
\begin{equation}
 f(z)=\frac{\sin\theta_0}{\sqrt2}\big(\cos\phi-\sin\phi\cos\theta_1\big)
 \label{eq:two-qubit-feature}
\end{equation}
is the sum of two products, each of which is the feature of a Clifford reservoir. The circuit without the $T$ gate returns $\cos\phi\sin\theta_0$, and the circuit with the $T$ gate replaced by the phase gate $\hat S$ returns $-\sin\phi\sin\theta_0\cos\theta_1$. The doped reservoir thus delivers through one measured string a fixed combination of two features that two Clifford reservoirs deliver separately.

The same argument applies to any reservoir. Expanded in Pauli strings, $\hat U^\dagger\hat Z_{\mathcal B}\hat U$ gives by Eq.~\eqref{eq:product-general} a sum of products of $1$, $\cos\theta_i$ and $\sin\theta_i$. The encoding fixes these functions~\cite{schuld2021effect}, and the unitary determines only their coefficients, namely a single product per measured string for a Clifford reservoir and at most $2^{n_T}$ products for $n_T$ transverse encounters with $T$ gates, whose evaluation cost grows exponentially with $n_T$~\cite{bravyi2016}.

In the expectation-value mode a head supplied with the individual products of Eq.~\eqref{eq:doped} absorbs the coefficients $c_k$ and reproduces the doped output, although on a device these products may correspond to noncommuting observables and require several measurement bases or Clifford reservoirs, as in the two-qubit example. In the single-shot mode doping changes only the latent means that fix $q_U(s)$, and not the parities received by the head. A doped reservoir can thus change the reachable distributions at a fixed readout, while it uses the same encoding functions.

\section{Random brickworks of generic gates}
\label{sec:haar}
\begin{figure}[t]
\includegraphics[width=\columnwidth]{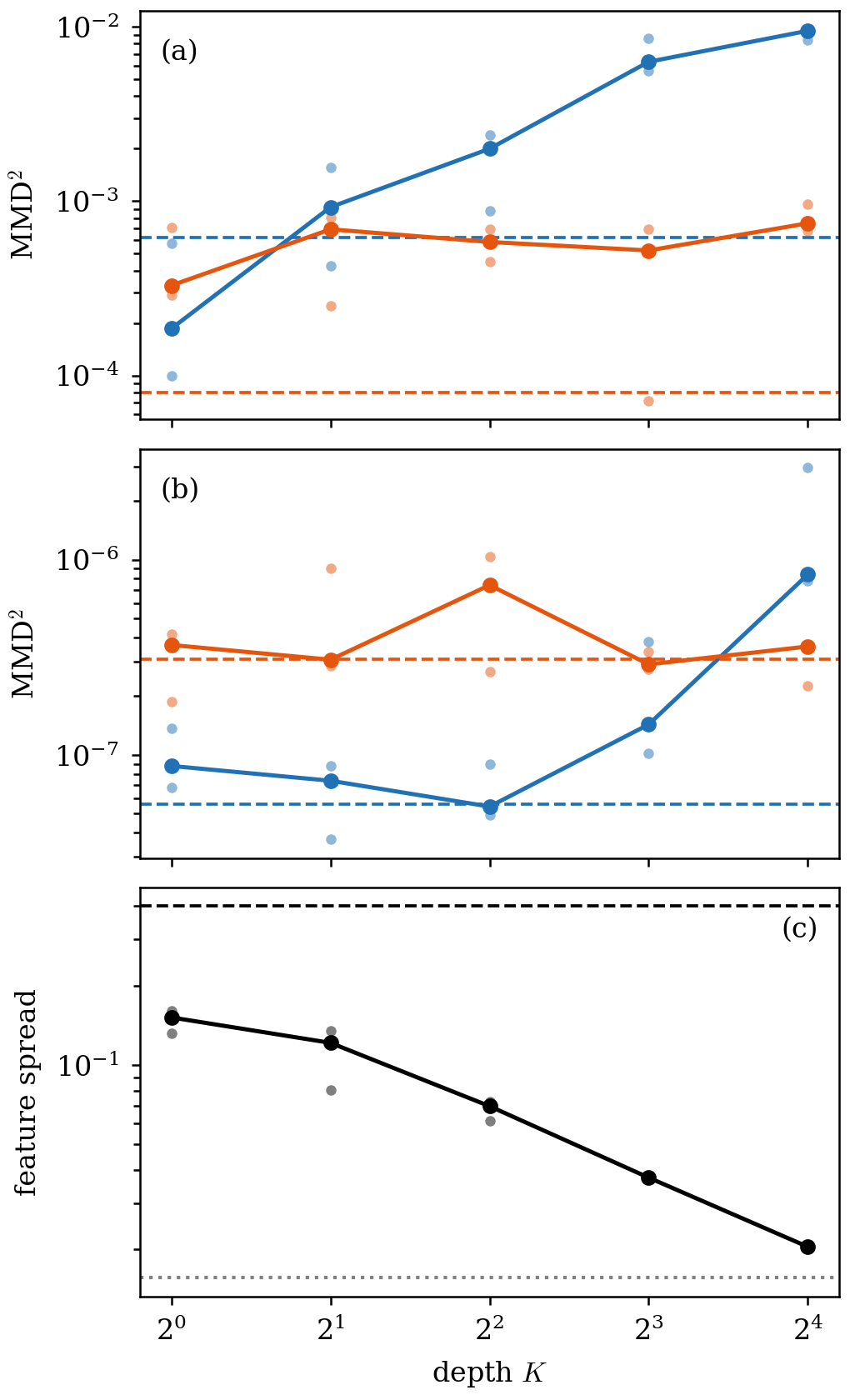}
\caption{Squared maximum mean discrepancy of Eq.~\eqref{eq:mmd} as a function of the depth $K$ of a brickwork of Haar-random two-qubit gates, for (a) the five-peak mixture and (b) the Gamma distribution, in the expectation-value mode (blue) and the single-shot mode (orange). Small points show three random brickworks for each depth and lines connect their medians. Dashed lines give the reservoir without gates, $\hat U_1$, in the same colors. (c) Spread $\Delta f$ of the features over the latent variable, Eq.~\eqref{eq:spread}, with the value without gates (dashed) and the Haar reference scale $1/\sqrt{2^N+1}$ from Eq.~\eqref{eq:haar-variance} (dotted). In all panels $N=12$ and $M=298$.}
\label{fig:haar}
\end{figure}

In the following, we consider reservoirs which are not Clifford circuits, namely brickworks in which every two-qubit gate is an independent Haar-random unitary, and we vary their depth $K$. With increasing $K$ such a brickwork approximates a Haar-random unitary acting on all $N$ qubits. Its features are no longer single products, and for this reason we compute them, together with the bitstring weights, directly from the state vector. Figure~\ref{fig:haar} shows the resulting discrepancies for the one-dimensional targets, with three random brickworks for each depth. In the expectation-value mode the generator deteriorates with growing depth. For the five-peak mixture the discrepancy increases by more than an order of magnitude between $K=1$ and $K=16$, and only brickworks with $K\le2$, among them all three with $K=1$, outperform the reservoir without gates. For the Gamma distribution the discrepancy begins to increase at $K=8$. In the single-shot mode, on the other hand, the discrepancy scatters from one brickwork to another without any systematic dependence on $K$, and the reservoir without gates is among the best ones for the five-peak mixture and lies in the middle for the Gamma distribution.

To compare the raw feature amplitudes at different depths, we recall the concentration of measure for Haar-random unitaries. If $\hat U$ is drawn uniformly from the unitary group of $N$ qubits, the expectation value of a traceless Pauli string $\hat P$ in a fixed state $\ket\psi$ is a random variable with zero mean and variance
\begin{equation}
 \E_{\hat U}\Big[\bra\psi\hat U^\dagger\hat P\hat U\ket\psi^2\Big]=\frac{1}{2^N+1},
 \label{eq:haar-variance}
\end{equation}
which follows from the second moment of the Haar measure and underlies barren plateaus~\cite{mcclean2018barren} and the concentration of quantum kernels~\cite{thanasilp2024exponential}. Typical expectation values are thus of order $2^{-N/2}$. In our protocol, however, a single reservoir acts on all encoded inputs, and we measure the variation of the features over $z$ at fixed $\hat U$ by the spread
\begin{equation}
 \Delta f=\frac1M\sum_{1\le|\mathcal B|\le3}\bigg[\int f_{\mathcal B}(z)^2\,dz-\bigg(\int f_{\mathcal B}(z)\,dz\bigg)^2\bigg]^{1/2},
 \label{eq:spread}
\end{equation}
i.e., the standard deviation of each feature over $z$, averaged over the $M$ measured strings. Without gates the features are products of cosines and $\Delta f$ is of order one. For a Haar-random $\hat U$, with $h=2^N$ and $\bar\rho=\int\ket{\psi(z)}\bra{\psi(z)}\,dz$, every measured feature has
\begin{equation}
 \E_{\hat U}\operatorname{Var}_z[f_{\mathcal B}(z)]
 =\frac{h(1-\Tr\bar\rho^2)}{h^2-1},
 \label{eq:haar-latent-variance}
\end{equation}
which follows by subtracting $\E_{\hat U}[\bar f_{\mathcal B}^{\,2}]=(h\Tr\bar\rho^2-1)/(h^2-1)$ from Eq.~\eqref{eq:haar-variance}. For our encoding $\Tr\bar\rho^2$ is of order $2^{-N}$, and Eq.~\eqref{eq:haar-latent-variance} equals $1/(2^N+1)$ up to corrections of relative order $2^{-N}$. Since $\Delta f$ averages standard deviations rather than variances, $1/\sqrt{2^N+1}$ is a reference scale for $\Delta f$ rather than its exact value.

The spread in Fig.~\ref{fig:haar}(c) decreases with $K$ and comes close to this scale at $K=16$, where the expectation-value fits are worst. The small amplitudes alone do not explain the deterioration, since we whiten the features before training, so that the head compensates any rescaling, and all $M$ directions are retained for every Haar brickwork. With growing depth, however, the features also change as functions of $z$, and the training can end in different minima, and we do not separate these two effects. On a device the small amplitudes are a problem of their own, since estimating an expectation value of order $2^{-N/2}$ to a fixed relative precision requires of order $2^N$ shots, as for concentrated quantum kernels~\cite{thanasilp2024exponential}.

The single-shot head receives parities $\pm1$ at any depth, and the reservoir enters only through the weights $q_U(s)$, whose Haar average is exactly uniform, although the weights of a particular brickwork need not be. Correspondingly, the single-shot fits show no systematic trend with depth.

\FloatBarrier
\section{Discussion and conclusions}
\label{sec:conclusions}
We proposed a generative quantum extreme learning machine with a fixed reservoir and a trained affine head. For Clifford reservoirs, Pauli propagation gives analytic features and bitstring weights for the expectation-value and single-shot modes.

For the targets studied, the reservoir without gates is competitive with the entangling Clifford circuits. Shallow Haar brickworks improve some expectation-value fits, but deeper ones worsen them. The decrease in raw feature amplitudes cannot alone explain this deterioration, since the head can rescale its inputs. Doping with $T$ gates mixes the encoding functions without adding new ones. The comparison of Clifford and near-Haar entanglers in our earlier Born-machine study~\cite{plodzien2026qbm} thus extends here to a generator with no entangler.

In the considered examples the quantum reservoir gives no advantage over its classical control. A head reading several shots at the same latent point would keep the correlations lost in the single-shot law, and it is an open question whether the reservoir then plays a larger role.

\appendix
\section{Bitstring weights and the Walsh transform}
\label{app:qu}
The single-shot generator depends on the device only through the distribution $q_U(s)$ of the measured bitstring averaged over the latent variable, Eq.~\eqref{eq:discrete}. Any function on the $2^N$ bitstrings $s\in\{\pm1\}^N$ can be expanded in the parities $s_{\mathcal B}$, which play the role of Fourier modes on the hypercube and satisfy the completeness relation
\begin{equation}
 \sum_{\mathcal B}s_{\mathcal B}\,s'_{\mathcal B}=2^N\delta_{s,s'},
 \label{eq:app-complete}
\end{equation}
with the sum over all subsets of qubits. This expansion is the Walsh transform. For a probability distribution $q(s)$, Eq.~\eqref{eq:app-complete} gives
\begin{equation}
 q(s)=\frac1{2^N}\sum_{\mathcal B}s_{\mathcal B}\,\E[s_{\mathcal B}],\qquad
 \E[s_{\mathcal B}]=\sum_{s'}q(s')\,s'_{\mathcal B},
 \label{eq:app-walsh-general}
\end{equation}
so that a distribution of bitstrings is fixed by the averages of all its parities. For the measured bitstring, the average of $s_{\mathcal B}$ is the expectation value of the string $\hat Z_{\mathcal B}$, which at fixed $z$ is the feature $f_{\mathcal B}(z)$. This is why the Born probability in Eq.~\eqref{eq:born} has the form of a Walsh expansion with the features as coefficients. Averaging over the latent variable replaces each feature by its latent mean
\begin{equation}
 \bar f_{\mathcal B}=\int_{[0,1]^D}f_{\mathcal B}(z)\,dz,
 \label{eq:app-mean}
\end{equation}
and the bitstring weights become
\begin{equation}
 q_U(s)=\frac1{2^N}\sum_{\mathcal B}s_{\mathcal B}\,\bar f_{\mathcal B}.
 \label{eq:app-walsh}
\end{equation}
The $2^N$ latent means, each obtained by Pauli propagation, are thus converted into the $2^N$ bitstring weights by one fast Walsh--Hadamard transform at a cost $O(N2^N)$.

For a Clifford reservoir every $f_{\mathcal B}$ is, by Eq.~\eqref{eq:product}, a product of the factors $\cos\theta_i$ and $\sin\theta_i$, and the factors of different latent coordinates are integrated independently. Writing the factors of one coordinate $z_v$ through $e^{\pm\mathrm i\theta_i}$, their product becomes a sum of terms $e^{\mathrm i\pi kz_v}$ with frequencies
\begin{equation}
 k=\sum_i\sigma_i2^{t_i},\qquad\sigma_i=\pm1,
 \label{eq:app-freq}
\end{equation}
where the sum runs over the qubits of this coordinate with a nontrivial factor. Every such $k$ is a nonzero integer, because the largest power of two in the sum exceeds the sum of all smaller ones. A product of cosines is therefore a sum of terms $\cos(\pi kz_v)$, and a product with an odd number of sines a sum of terms $\sin(\pi kz_v)$, which integrate to
\begin{equation}
 \int_0^1\cos(\pi kz)\,dz=0,\qquad
 \int_0^1\sin(\pi kz)\,dz=\frac{1-(-1)^k}{\pi k}.
 \label{eq:app-int}
\end{equation}
Only odd $k$ contribute, and $k$ is odd exactly when the qubit with $t_i=0$ carries a factor. A product therefore has a nonzero mean only if, for every coordinate it depends on, it contains an odd number of sines and a factor on this qubit. Without a reservoir every $f_{\mathcal B}$ with $\mathcal B\neq\emptyset$ is a product of cosines, so that $\bar f_{\mathcal B}=0$, only the empty set contributes to Eq.~\eqref{eq:app-walsh}, and
\begin{equation}
 q_I(s)=2^{-N}
 \label{eq:app-uniform}
\end{equation}
for every bitstring. The conditional distribution $p_I(s\,|\,z)$ is not uniform, but only its latent average enters the single-shot generator.

\section{Numerical methods}
\label{app:methods}

\textit{Targets.} Every target is a fixed distribution $p_j$ on the $2^{12}$ grid nodes, obtained by evaluating a continuous density at the nodes and normalizing. The five-peak mixture on the bin index $j=0,\dots,4095$ reads
\begin{equation}
 p_j\propto\sum_{k=1}^{5}w_k\exp\!\Big[-\frac{(j-\mu_k)^2}{2\sigma^2}\Big],
 \label{eq:app-fivepeak}
\end{equation}
with $\mu_k=(k-\tfrac12)2^N/5$, $\sigma=2^N/20$ and weights $w=(1.593,0.990,1.745,1.455,0.370)$, given here rounded and drawn uniformly from $[0.2,2]$ with NumPy's \texttt{default\_rng} and seed $42$, as in Ref.~\cite{plodzien2026qbm}. The Gamma target reads
\begin{equation}
 p_j\propto y_j\,e^{-y_j},\qquad y_j=\frac{12j}{4095}.
 \label{eq:app-gamma}
\end{equation}
The two-dimensional targets are defined on $64\times64$ equally spaced nodes in $[-3,3]^2$. The four lobes are Gaussians of width $0.5$ centered at $(\pm1.5,\pm1.5)$, and the rings, the stripes and the cross are sums of Gaussian ridges,
\begin{equation}
 p(\vec y)\propto\sum_c\exp\!\Big[-\frac{d_c(\vec y)^2}{2\delta^2}\Big],
 \label{eq:app-ridge}
\end{equation}
where $d_c(\vec y)$ is the distance to the curve $C_c$, namely two circles of radii $1.0$ and $2.3$ at $\delta=0.26$, three lines $y_1-y_2\in\{-2.5,0,2.5\}$ at $\delta=0.35$, and the diagonals $y_1=\pm y_2$ at $\delta=0.28$. The two moons follow the standard two-moons geometry of Ref.~\cite{plodzien2026qbm}, as a single ridge at $\delta=0.30$ around the nearer of two semicircles of radius $2$, which we center at $(-1,-0.5)$ and $(1,0.5)$.

\textit{Circuits and training.} The random brickwork $\hat U_3$ is generated with the random-number seed $1$, and the nine further brickworks with the seeds $0$ and $2$ to $9$. In the two-dimensional expectation-value mode only the seeds $0$, $1$ and $2$ were used. For the random brickworks of Haar-random two-qubit gates of Sec.~\ref{sec:haar}, we use the seeds $0$, $1$ and $2$ at each depth and compute the features and the weights from the full state vector, the latter with an $8192$-node Gauss--Legendre quadrature over the latent variable, which reproduces the analytic Clifford weights at $N=12$ to $4.4\times10^{-13}$. The training and the evaluation are the same as above. In 2D, qubits $0$--$5$ encode $z_1$ and qubits $6$--$11$ encode $z_2$. For Clifford reservoirs the features and the weights follow from Eqs.~\eqref{eq:product} and \eqref{eq:discrete}, and both were checked against a dense statevector simulation. We whiten the features, removing redundant directions, and map the 1D coordinates affinely to $[-5,5]$. We then minimize Eqs.~\eqref{eq:w2} and \eqref{eq:sliced} with L-BFGS-B, with at most $2200$ (1D) and $1600$ (2D) iterations, starting from three random heads matched to the target mean and covariance and keeping the one with the lowest training loss. The losses reached from different starts of the same reservoir can differ as much as those of different reservoirs, so that the ranks among reservoirs quoted in the main text indicate tendencies rather than a significant ordering. Of the $193$ selected heads, $133$ for Clifford reservoirs and $60$ for Haar brickworks, seven Clifford heads for two-dimensional targets stopped at the iteration limit, and all others met the stopping criterion of the solver, which does not certify a global minimum. In the expectation-value mode we train on $2^{15}$ latent points and evaluate on a separate set of $2^{17}$ points. In the single-shot mode we sum over all $4096$ bitstrings, with the analytic weights for Clifford reservoirs and the quadrature weights for the Haar brickworks. When a generated point is assigned to the nearest grid node, coordinates outside the grid are moved to its boundary, and a point lying exactly halfway between two nodes is assigned to the upper one. The probability moved to the boundary in this way is the largest, about $0.04$, for the single-shot identity model on the three stripes, and the values for all settings can be found in the released data.

\textit{Evaluation.} We evaluate Eq.~\eqref{eq:mmd} in the training coordinates. We also evaluate the total variation distance $\frac12\sum_j|\nu_j-p_j|$ after assignment to grid nodes. For the comparison of Sec.~\ref{sec:2d}, the single-shot heads trained with $\hat U_3$ are evaluated, without retraining, under the uniform weights of $\hat U_1$.

\section*{Data and code availability}
The code, the trained heads and the data needed to reproduce all figures are available at \url{https://github.com/MarcinPlodzien/generative-QELM}.

\bibliography{references}
\end{document}